\documentclass[aps,prl,longbibliography,twocolumn,superscriptaddress]{revtex4-1}
\usepackage{graphicx}
\usepackage{soul}
\usepackage{color}

\begin{document}


\title{How moving cracks in brittle solids choose their path}


\author{Lital Rozen-Levy}
\affiliation{The Racah Institute of Physics, The Hebrew University of Jerusalem, Givat Ram,
	Jerusalem Israel}
\author{John M. Kolinski}
\affiliation{\'{E}cole Polytechnique F\'{e}d\'{e}rale de Lausanne, 1015 Lausanne, Switzerland}
\author{Gil Cohen}
\author{Jay Fineberg}
\affiliation{The Racah Institute of Physics, The Hebrew University of Jerusalem, Givat Ram,
	Jerusalem Israel}


\date{\today}

\begin{abstract}
While we fundamentally understand the dynamics of ‘simple’ cracks propagating in brittle solids within perfect (homogeneous) materials, we do not understand how paths of moving cracks are determined.  We experimentally study strongly perturbed cracks that propagate between 10-95\% of their limiting velocity within a brittle material. These cracks are deflected by either interaction with sparsely implanted defects or via an intrinsic oscillatory instability in defect-free media. Dense, high-speed measurements of the strain fields surrounding the crack tips reveal that crack paths are governed by the direction of maximal strain energy density. This fundamentally important result may be utilized to either direct or guide running cracks.

\end{abstract}


\maketitle


So long as a crack propagates along a straight path, we have an excellent understanding of its motion \cite{Freund1998,bouchbinder_dynamics_2014, goldman_acquisition_2010, livne_near-tip_2010} within perfect (homogeneous) materials. Cracks, however, can change direction. How is the direction selected by a propagating crack determined?  In ideal brittle materials, the elastic fields at the tips of moving cracks are singular \cite{Freund1998}. For such cracks empirical criteria for path selection include the  “principle of local symmetry," \cite{goldstein_brittle_1974} whereby a crack will rotate so as to negate any singular shear (`Mode II' component) at its tip, and the “principle of maximal strain energy density”, whereby a crack will choose the direction that maximizes the strain energy dissipated at its tip \cite{slepyan_principle_1993}. 2D analysis \cite{amestoy_crack_1992} and experiments \cite{erdogan_crack_1963} suggest that both criteria for path selection  are approximately equivalent for quasi-static cracks, whose velocities are a small fraction of their limiting velocity, the Rayleigh wave speed, $c_R$. 3D calculations \cite{hodgdon_derivation_1993} suggest that quasi-static cracks, beyond a transitional region set by the microscopic scales surrounding a crack's tip, indeed respect the principle of local symmetry. Energy considerations \cite{Francfort_1998} have also been used to identify quasi-static crack paths.

The fundamental question, however, of what determines a {\it dynamic} crack’s path has long been open \cite{fineberg_instability_1999, chen_instability_2017}. Dynamic cracks propagate at a finite fraction of $c_R$. While the principle of local symmetry predicts that the singular shear (Mode II) component must always be zero at crack tips, experiments reveal \cite{boue_origin_2015} that a non-zero Mode II component can actually be sustained at the tips of {\em dynamic} cracks propagating along {\it straight} trajectories. This is consistent with the linear stability analysis of a straight dynamic crack’s path to singular shear stresses; straight cracks are stable until losing their stability to infinitesimal perturbations  \cite{bouchbinder_dynamic_2009}  at extreme velocities. Experimentally, this instability corresponds to intrinsic path oscillations \cite{chen_instability_2017, livne_oscillations_2007}. The path selected by these oscillations is not a-priori known.

These oscillations occur in perfect, `homogeneous' materials. When materials have defects, slow cracks interact with them, and the crack front can pin or deform\cite{vasoya_finite_2016, dalmas_crack_2009, kolvin_topological_2018, santucci_fracture_2010}. Interactions with defects  may also deflect \cite{lee_sideways_2019, steinhardt_rules_2019} or even arrest a crack. How such defects will affect the paths of dynamic cracks is an open question. 

\begin{figure}
\includegraphics[width=\columnwidth]{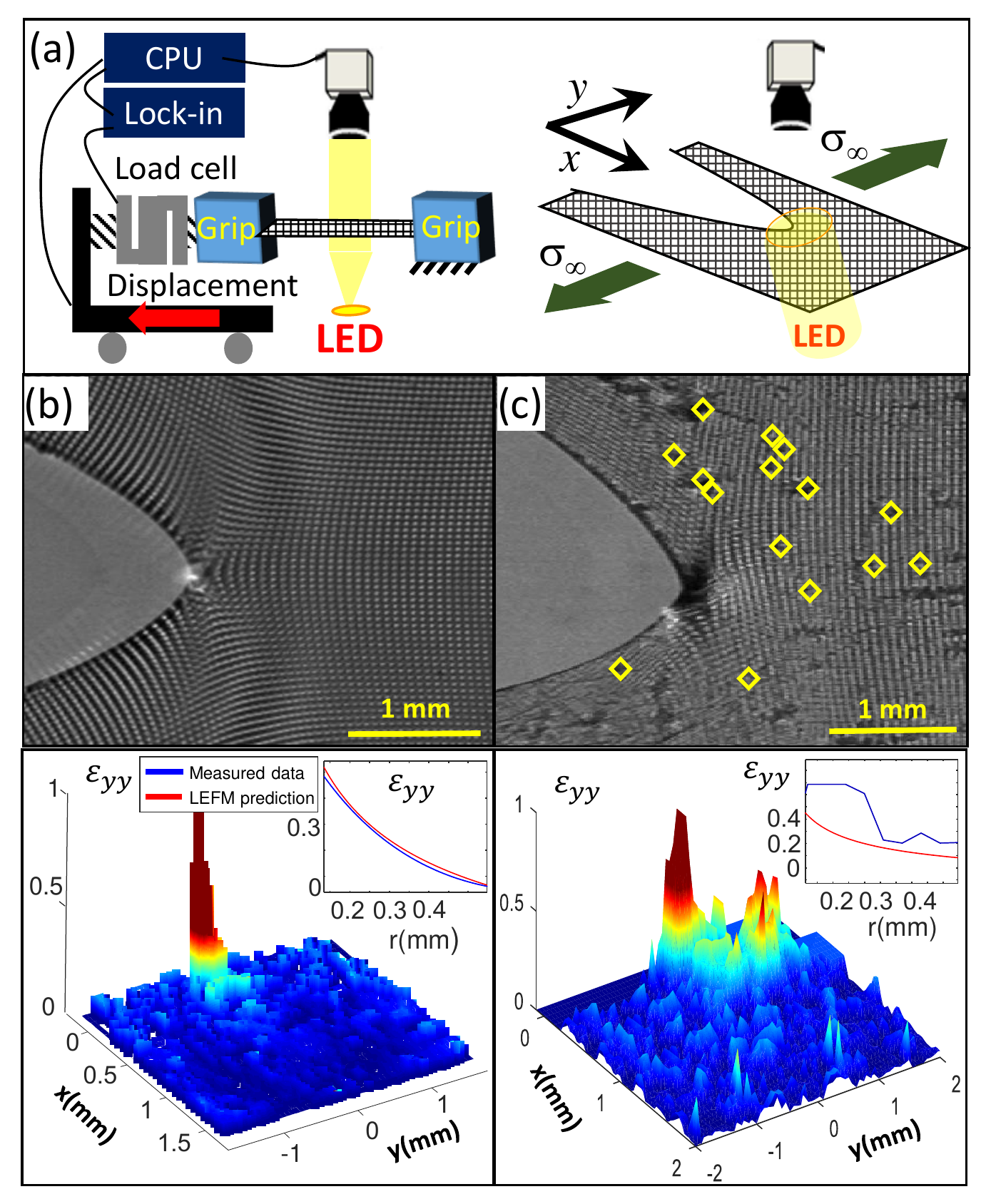}%
\caption{Experimental system. (a) Brittle polyacrylamide gel sheets are loaded in tension, and dynamic cracks are initiated at sheet edges. Collimated LED beams produced shadowgraph images of shallow grids embossed on gel surfaces. Grid deformations provide instantaneous displacement and strain fields near crack tips. (b) (top) A crack ($v\sim 0.25 c_R$) within a pure gel sample with no embedded particles. (bottom) The measured $\varepsilon_{yy}$ strain component has the  $r^{-1/2}$ scaling (inset) predicted by LEFM  of  $\varepsilon_{yy}$ along $y=0$. (c) (top) A crack ($v\sim 0.3 c_R$) approaches embedded polyamide spheres (highlighted by diamonds). (bottom)  Nearby particles strongly affect the $\varepsilon_{yy}$ field (see also \cite{supplementary_material}). (inset) $\varepsilon_{yy}$ along the path selected by the crack (blue) strongly  deviates from  LEFM predictions (red). \label{fig1}
}
\end{figure}

Here, we experimentally probe the dynamic path selection of cracks resulting from both dilute rigidly embedded inclusions and the intrinsic oscillatory instability in perfect (homogeneous) brittle isotropic materials. 

We visualize fully dynamic propagation by using polyacrylamide \cite{goldman_acquisition_2010, livne_universality_2005} hydrogels. These compliant materials, in which $c_R$ is slowed to $\sim$5.3m/s, are representative of the broad class of materials that undergo brittle failure; the fields at their tips are singular \cite{bouchbinder_dynamics_2014, livne_near-tip_2010, bouchbinder_dynamic_2009, goldman_intrinsic_2012} and they have been used to verify predictions of fracture mechanics for cracks propagating at all velocities.

Here, we use polyacrylamide hydrogels composed of a total monomer concentration of 13.7\% (wt) with a 2.7\% (wt) cross-linker as in  \cite{bouchbinder_dynamics_2014, livne_near-tip_2010, bouchbinder_dynamic_2009, goldman_intrinsic_2012}. Details of our materials and methods are provided in \cite{supplementary_material}. We define the coordinate directions in the unstressed frame of reference, such that $y$ is aligned with the applied loading, and $x$ is perpendicular to this direction. The vertical ($y$-direction) boundaries of the gel samples were sandwiched between two grips that were rigidly displaced to produce desired strains. The magnitude of the applied strain governed the mean crack velocities in the measurement area. We used `strip' samples ($x \times y$  dimensions of  $60 \times 40$ mm) to study crack-particle interactions. Samples of size $200 \times 200$mm were used for the oscillatory instability studies. Strip samples were sparsely seeded with $50\mu m$ diameter rigid particles that adhere tightly to the surrounding gel (see \cite{supplementary_material}).  The center $10 \times 6$mm sections were illuminated, and imaged with a fast camera (IDT Y7 S3) at 8000 fps, as shown schematically in Fig.~{\ref{fig1}}(a). Each of the camera's 2000 x 1000 pixels was mapped to $6\mu m \times 6\mu m$.  Applied strains varied from 10 - 25\% in strips and were about 6\% in oscillatory experiments. At desired strains, cracks were triggered by cutting the middle of sample edges at $x=0$. 
	
Deformation fields were determined by measuring the distortions of a $2\mu m$ deep reference grid embossed on one surface of each sample. In the unstressed `reference' frame, grid dimensions were $60 \mu m$ along each side. Deformations were measured by comparing the distances between adjacent grid points in the laboratory frame to the unstressed grid. This enabled high precision deformation measurements in the reference frame, despite the strong rotations and high strains near crack tips that would challenge cross-correlation methods. Grid point locations were determined to 1/3 pixel.  We calculated the deformation gradient tensor via finite differences of neighboring grid displacements. Corresponding stresses were calculated using Neo-Hookean elasticity \cite{Knowles.83,bouchbinder_dynamics_2014} and plane-stress conditions were assumed. Fig.~{\ref{fig1}}(b,c) presents typical results for homogeneous and particle-embedded samples near dynamic crack tips. 

\begin{figure}
	\includegraphics[width=\columnwidth]{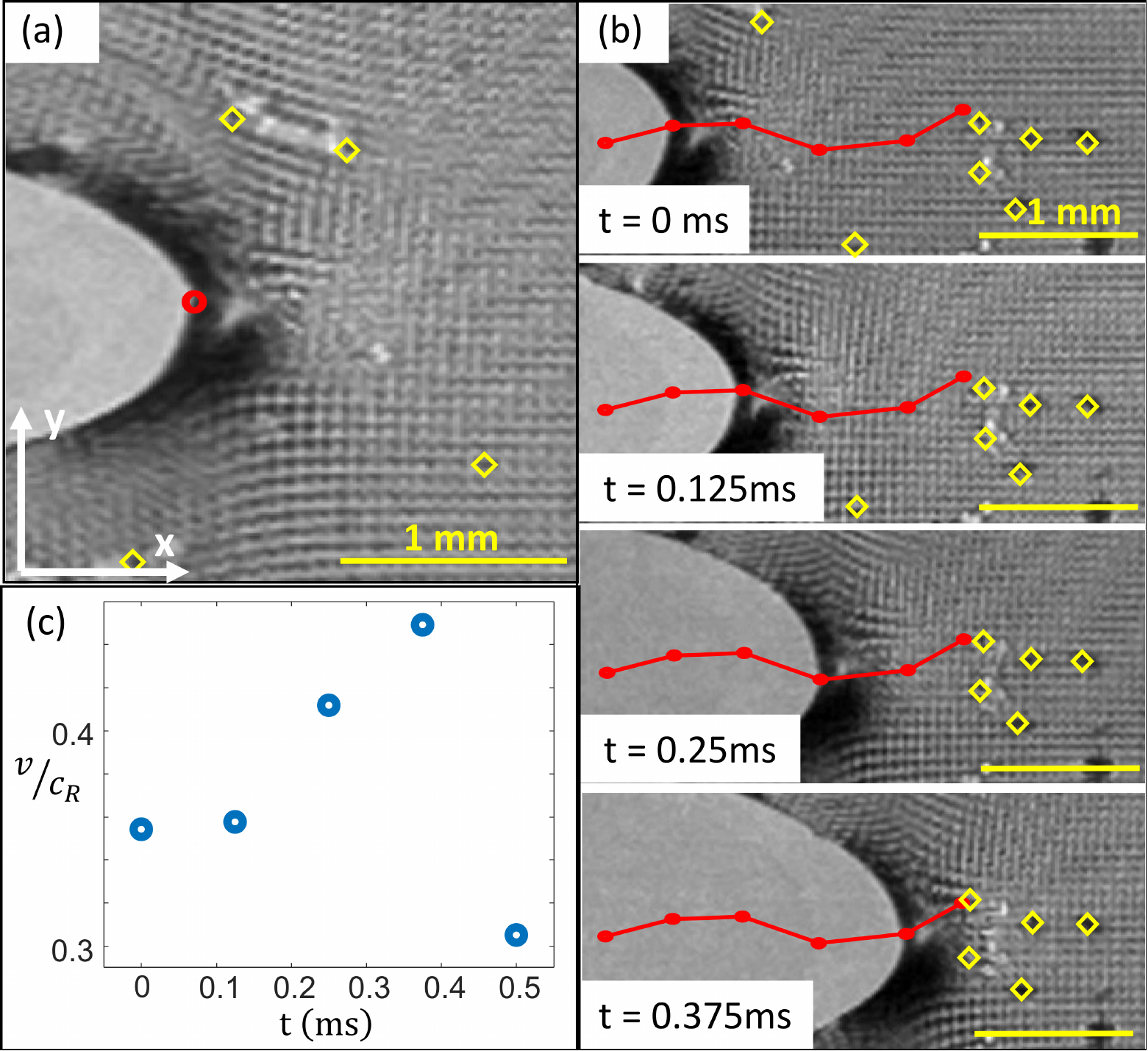}
	\caption{Crack tip path deviation and acceleration upon encountering particles. (a) Interactions with embedded particles cause crack tip opening displacements (CTOD) to become non-parabolic and force cracks to deviate significantly from straight paths. We define a crack tip as the leading point in the CTOD, as denoted by the red dot. (b) A time series of images separated by 0.125msec, as a crack rapidly evolves while interacting with embedded particles (yellow diamonds). Its resulting path, denoted by the red line connecting successive crack tips (red dots), becomes tortuous. (c) Corresponding local tip velocities, normalized by $c_R=5.3$ m/s, as a function of time. Such particles may even induce crack arrest  \cite{supplementary_material}. \label{fig2}}
\end{figure}

Paths of dynamic cracks with velocities $0.1< v/c_R<0.8$ are often tortuous due to interactions with particles. Nearby particles alter or even destroy the singular strain fields predicted by fracture mechanics. When no particles are embedded, the predicted singular fields are obtained {\cite{boue_origin_2015, livne_breakdown_2008}}. Since gels are constrained to zero displacement at rigid particle boundaries \cite{supplementary_material}, strain fields deviate strongly from the singular, ‘K-dominant’ ($K/r^{1/2}$) scaling predicted by LEFM and its nonlinear extensions \cite{bouchbinder_weakly_2008, bouchbinder_dynamics_2014, livne_breakdown_2008, ronsin_crack_2014, long_crack_2015}, as shown in Fig.~\ref{fig1}c. 

Why do rigid microscopic inclusions induce such large perturbations? A rigid inclusion imposes an additional boundary condition, negligible strain, at its boundaries. Studies of the interactions between cracks and inclusions go back to the beginnings of fracture mechanics \cite{atkinson_interaction_1972, erdogan_interaction_1974}; calculated stress intensity factors for static cracks can be significantly either enhanced or reduced by interactions with an inclusion. These predictions are consistent with our experimental observations of dynamic cracks, where the presence of even a single rigid inclusion will produce a huge perturbation at any location where unperturbed stresses are large, such as near a crack tip. Even if an inclusion is minute in size, the resultant pinning of the strain field at the inclusion will entirely alter the magnitude and symmetry of the stresses surrounding a crack’s tip. This is clearly exemplified in the example of crack arrest presented in \cite{supplementary_material}.

Interactions with embedded particles can, therefore, profoundly affect crack propagation. This is readily observed in even a single image, as the parabolic crack tip opening displacement (CTOD) predicted from LEFM is distorted and asymmetric (Fig. \ref{fig2}(a)). In the strip geometry, cracks in homogeneous media propagate at steady velocities  \cite{fineberg_instability_1999, goldman_acquisition_2010}. Upon encountering particles, cracks can significantly change their velocity and direction within microseconds, as seen in Fig. \ref{fig2}(b,c).

By calculating the local deformation tensor, we can evaluate path selection criteria. All such calculations are both carried out and presented in the material reference frame. These measurements enable us to calculate the strain energy density (SED) at each grid point (see \cite{supplementary_material} for details). Figure~\ref{fig3} presents a typical example in which a moving crack has changed its direction. In the figure we compare the direction predicted by the maximal strain energy density criterion with the crack's actual path. Presented is the map of the strain energy density at each grid point (the contribution of the background strain is included). Starting at the crack tip, for each $x$-position we note the $y$-location in which the strain energy density is maximal. The direction, determined by fitting a straight line to the energy density maxima ahead of the crack tip, is in excellent agreement with the propagation direction that leads to the subsequent crack tip location. 


\begin{figure}
\includegraphics[width=\columnwidth]{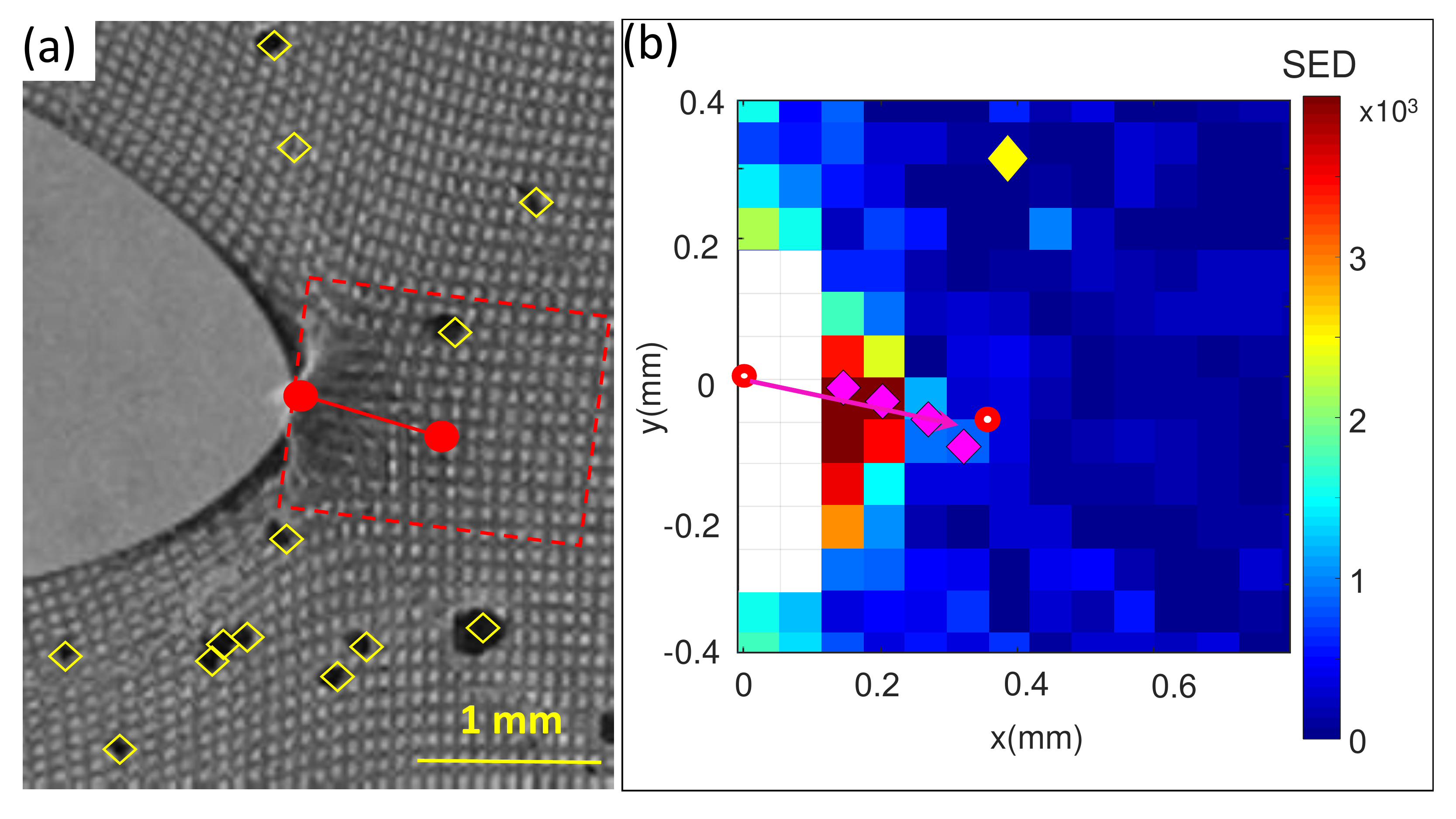}%
\caption{The SED-based prediction agrees well with observed propagation directions. (a) A typical crack whose trajectory is governed by crack-particle interactions. The trajectory is indicated by the red line. Particles are outlined by yellow diamonds. (b) The strain energy density (SED, in J/m$^3$) field ahead of the crack within the dashed red box in a). Magenta diamonds denote the maximal value for the SED for each value of $x$ in the reference frame. Current and future tip locations are indicated by red dots. Linear fits (magenta arrow) to the local SED maxima ahead of the tip agree well with the future crack tip location. Yellow diamond: a particle location.  \label{fig3}}
\end{figure}

\begin{figure}
\includegraphics[width=8.5cm]{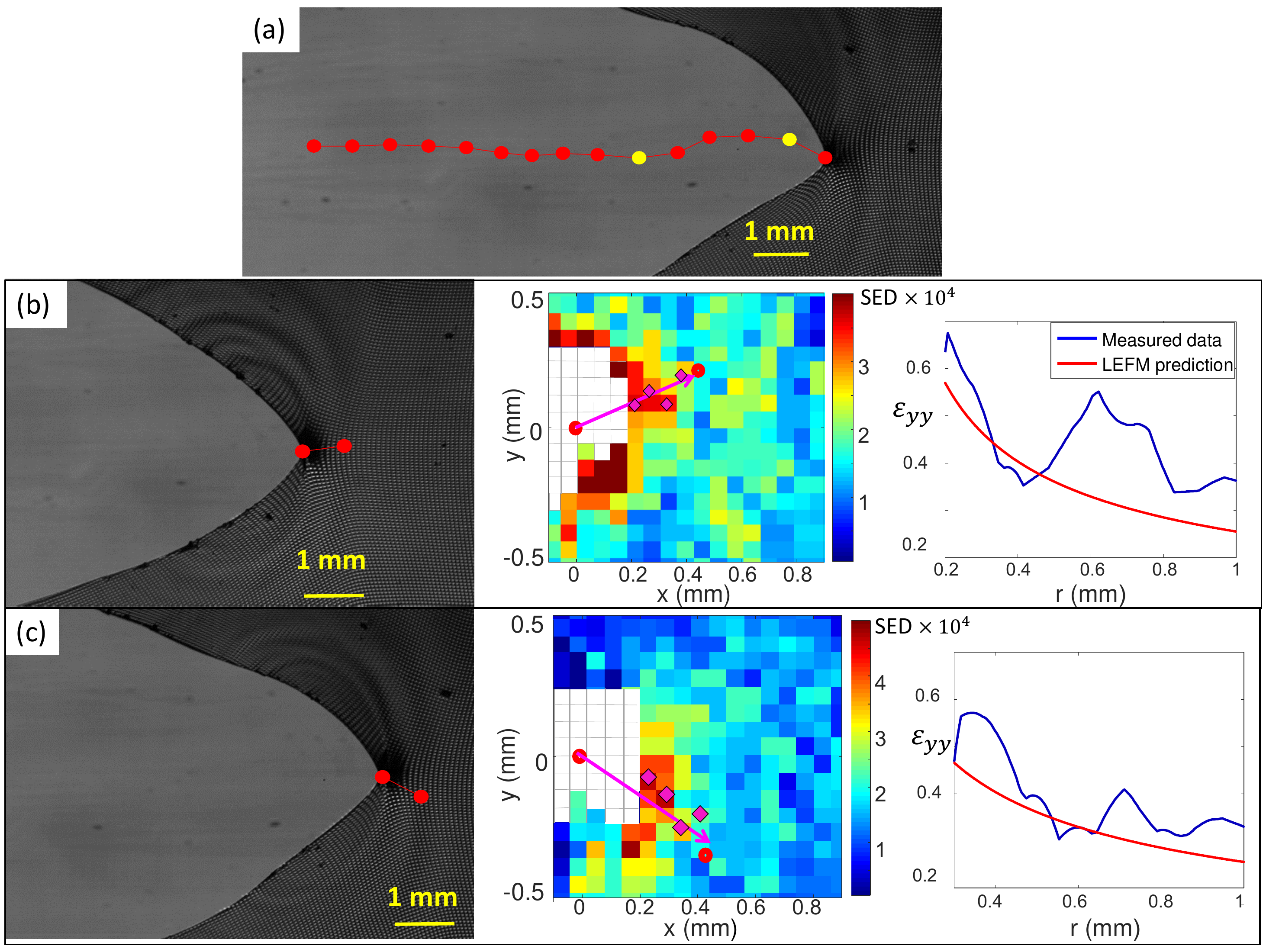}%
\caption{The maximum strain energy density criterion (SED) predicts the instantaneous direction of oscillatory cracks. (a) The trajectory of a spontaneously oscillating crack propagating at approximately $0.95c_R\sim 0.9c_s$ in a homogeneous gel at the onset of the oscillatory instability \cite{livne_oscillations_2007, goldman_intrinsic_2012}. The trajectory of successive crack tips leading to this instant is denoted in red. (b,c) At left, snapshots of the crack tip in the laboratory frame that are indicated by yellow dots in (a). Center panels present the respective SED fields, in J/m$^3$, in the material area ahead of each crack tip. As in Fig. \ref{fig3}b, diamonds denote the locations of the maximal SED value for each $x$. The line fitted to these maxima agrees well with the propagation direction denoted by the red line connecting the sequential tips (red dots).  Predicted (measured) angles were 30$^o$ (31$^o$) in b) and -40$^o$ (-36$^o$) in c). Right panels: $\varepsilon_{yy}$ (in blue) along the trajectory direction compared to the LEFM prediction (red).  Singular fields are highly perturbed by large amplitude waves excited by crack oscillations.   \label{fig4}}
\end{figure}

In Fig.~\ref{fig4} we consider an experiment where a straight crack loses stability to oscillations, while propagating through a particle-free medium.  When  $v>0.9c_s$, cracks undergo spontaneous oscillations \cite{chen_instability_2017, livne_oscillations_2007,  bouchbinder_dynamics_2014} that rapidly deflect their tips by large ($ > 30^o$) angles. In Fig.~\ref{fig4}b,c we analyze two typical instances. As we observed for cracks deflected by particles (Fig.~\ref{fig3}), the criterion of maximal SED again provides excellent predictions for the directions instantaneously selected by the crack. During these oscillations, stress fields at the crack tip are strongly perturbed, relative to those of a straight crack (Fig.~\ref{fig4}b,c - right). These strong perturbations are due to loss of symmetry accompanied by high-amplitude waves emitted by these rapidly oscillating cracks that transmit the crack's history to the crack tip vicinity. These perturbations to $\varepsilon_{yy}$ are roughly the size of those resulting from crack-particle interactions (see Fig.~\ref{fig1}c - inset).   

\begin{figure}
\includegraphics[width=\columnwidth]{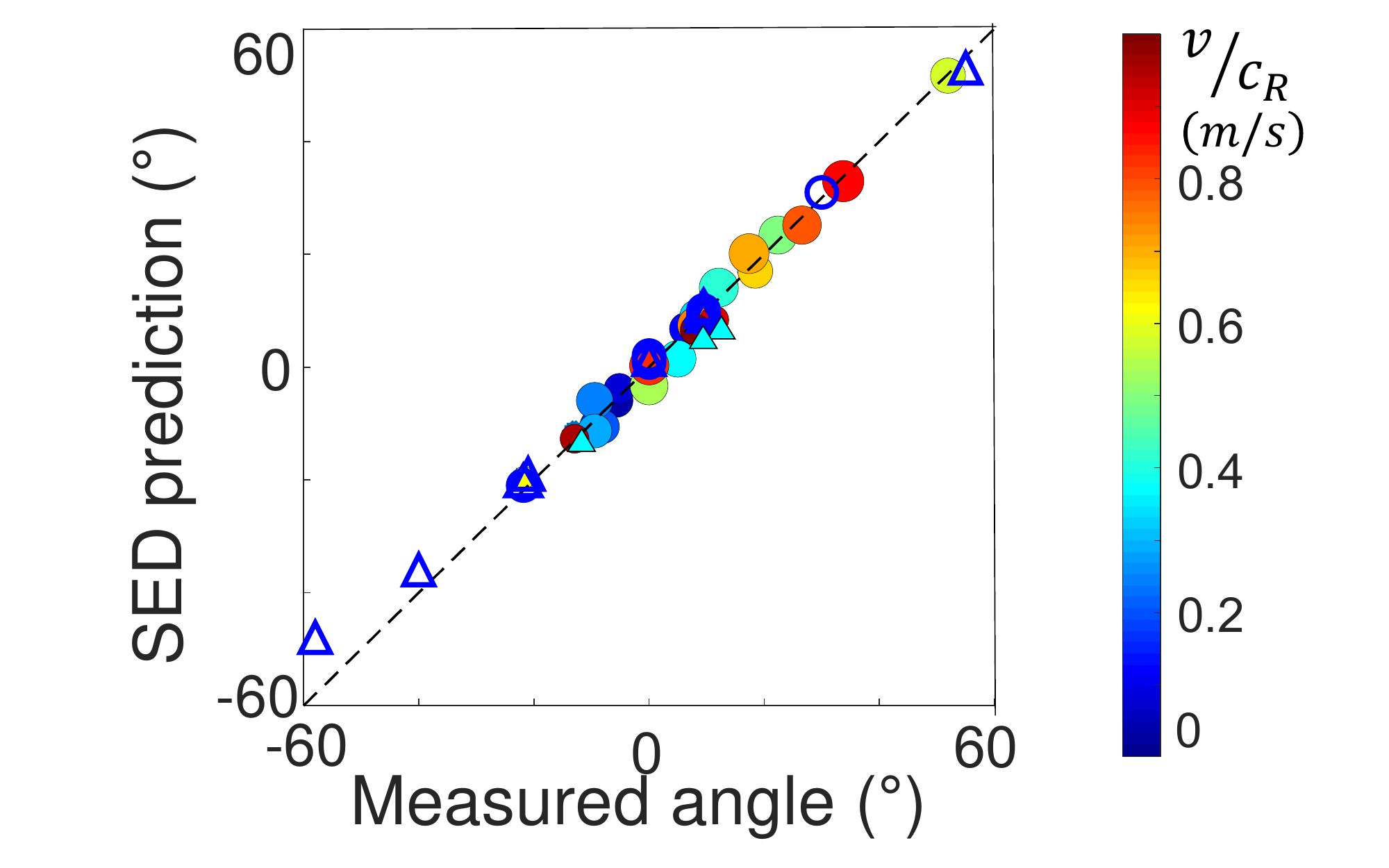}%
\caption{All deflected crack directions are accurately predicted by the maximum strain energy density (SED) condition. Filled circles: predicted angles for 21 crack-particle interactions with differing $v/c_R$ (colors) and particle ensembles (2 – 6 particles/mm$^2$). Open symbols are due to the oscillatory instability. Open (closed) triangles are sequential data from Fig.~\ref{fig4} (Fig.~\ref{fig2}). Horizontal (vertical ) error bars are the (1/3 of the) point sizes. Dashed line represents perfect agreement.  \label{fig5}}
\end{figure}

Are the examples in Figs.~\ref{fig3} and \ref{fig4} representative of dynamic crack path selection in general? In Fig.~\ref{fig5} we compare the predictions of the SED criterion to measured values from numerous experiments, where  $0.1<v/c_R  <0.95$ and selected propagation directions ranged over $\pm60^o$. The deflected angles corresponded to both crack-particle interactions as well as from the oscillatory instability. As the locations and numbers of particles within each frame are random, each crack-particle interaction experiment relates to an entirely different configuration of particles relative to both the crack tip and crack orientation.  We see that the maximal strain energy density criterion (Fig.~\ref{fig5}) agrees well with all of the measured paths. The accuracy of the predictions is therefore independent of the details of how crack deflections take place.  

The results in Fig.~\ref{fig5} clearly demonstrate that a dynamic crack’s path is determined by the SED criterion. It has long been known that, upon initiation, directions selected by static cracks in homogeneous media (when subjected to remote mixed-mode loading ) \cite{amestoy_crack_1992, erdogan_crack_1963} are consistent with both the SED and principle of local symmetry. Both criteria were originally conceived for cracks whose stresses are determined by a dominant leading-order singularity at their tip. Theoretically, both criteria are essentially indistinguishable when small perturbations are applied to the singular fields prescribed by LEFM \cite{amestoy_crack_1992}. 

What has fundamentally changed when material pinning locations are dispersed along a crack’s path or strong oscillations of a crack's tip occur? Insight is provided by the near-tip strain fields along the propagation direction (Figs.~\ref{fig1}(c) and \ref{fig4}), which show that the structure of the deformation fields ahead of the crack tip is no longer the ubiquitous singular strain fields characteristic of homogeneous materials. The near-tip stress fields in both cases are strongly varying in space and time, generally asymmetric and often effectively blunt the singularity of the fields within this region. These significant deviations from singularity apparently render the PLS inapplicable (see \cite{chen_instability_2017}), even as cracks continue to propagate and select their direction. In contrast, the SED criterion, when performed in the `intermediate' region in which the singular fields are {\it not} dominant, remains an excellent predictor of a crack’s local path. 

The characteristic parabolic form of crack tips away from the near-tip region suggests, however, that on average, the singular fields may be retained at spatial scales that are much larger than the spacing between inhomogeneities. At these scales, the observed {\it mean} path deviations become quite small. As we demonstrate in \cite{supplementary_material}, the direction of the axis of symmetry of the parabolic CTOD formed away from the crack tip (but still close enough that information from the crack tip can reach them in the 0.125msec intervals between frames) corresponds to the direction in which shear within the asymptotic region near the crack tip is zero. Therefore, in an `mesoscopic' sense, PLS may be restored. 

One might argue that the local path selection embodied in the SED criterion may be less relevant than a criterion (such as the PLS) that works at the mesoscale. Is predicting the erratic meandering of a crack's tip in response to localized perturbations necessary, if a crack's mean direction can be predicted?  In \cite{supplementary_material}, we demonstrate that a crack tip’s interaction with even a single isolated particle in the near field can be so significant that a crack can transition from rapid ($v=0.4c_R$) fracture to complete arrest. Moreover, we have seen that the SED is incredibly good at describing crack tip directions in unperturbed isotropic materials when cracks become unstable to the oscillatory instability at extreme velocities ($v>0.9 c_R$)  \cite{livne_oscillations_2007}. In both of these cases, crack dynamics at the near-tip scale are critical. While the far-field scales supply the energy flux to the crack tip needed for propagation, the local stress configuration does have global consequences, determining how, if, and when a crack will propagate. 

In short, this work provides clear experimental evidence that the maximum SED criterion clearly selects crack directions. This empirical observation both raises a theoretical challenge and provides a possible way to better understand how imperfections influence the fracture of heterogeneous or hybrid materials.  

J.F. L. L-R., and G. C. acknowledge the support of the Israel Science Foundation (grants 1523/15 and 840/19). J. M. K. and L. L-R. contributed equally to this work.


%

\end{document}